\documentclass[12pt]{article}
\usepackage{graphicx}
\begin{document}
\title{Calculation of Four Point Correlation Function of Logarithmic Conformal Field Theory Using AdS/CFT
Correspondence}
\author{S.Jabbari-Faruji    \footnote{e-mail: jabbari@mehr.sharif.ac.ir}  and
S.Rouhani
\footnote{e-mail: rouhani@ipm.ir}\\
\\
Department of Physics, Sharif University of Technology,\\ Tehran,
P.O.Box: 11365-9161, Iran}
\date{}
\maketitle
\begin{abstract}
 We use the correspondence between scalar field theory on AdS
and induced conformal field theory on its boundary to calculate
correlation functions of logarithmic conformal field theory in
arbitrary dimensions.Our calculations utilize the newly proposed
method of nilpotent weights.We derive expressions for the four
point function assuming a generic interaction term.
\\
\\
{\it keywords:conformal field theory,AdS/CFT  correspondence}
\end{abstract}
\section{Introduction}
            The relationship between field theories in (d+1)-dimensional anti-desitter space and d-dimensional
          conformal field theories was first suggested by Maldacena  and since then much work has been done
          on various aspects of this correspondence[2,3].
            This conjecture can be stated as follows,consider the action $S[\Phi]$ defined on $AdS_{d+1}$ and let
            $\Phi_{b}$
          be the value of $\Phi$ on the boundary:
          \begin{equation}\label{eq:a1}
                                \Phi|_{\partial AdS}=\Phi_{b}
                                \end{equation}

         So the partition function of the AdS theory subjected to this constraint is:
\begin{equation}
         Z_{AdS}[\Phi_{b}]=\int_{\Phi_{b}}
         D\Phi\exp(-S[\Phi])\end{equation}
          where the path integral is over configurations fulfilling
         (\ref{eq:a1}).

         The correspondence states that the partition function
         of AdS theory is the generating functional of the
         boundary conformal field theory.

               \begin{equation}\label{eq:a2}
               Z_{AdS}[\Phi_{b}]=\langle \exp(\int_{\partial AdS} dx \hat{O}\Phi_b)\rangle
                \end{equation}

         The function $\Phi_b$ is considered as a current which couples to  the
         scalar conformal operator $\hat{O}$ via  a coupling $\int_{\partial AdS}dx
         \hat{O}\Phi_b$.
         This is an elegant and useful result, since it gives a
         practical way for calculation of correlation functions of
         conformal field theory .
        However since 2 and 3-point function are fixed ( up to
         a constant) by conformal invariance , one is specially
         interested in  $n>3$.
         The conformal correlators have been studied for various
         cases, e.g.interacting massive scalar field theory [4] or interacting scalar spinor field
         theory [5].It is also interesting to find actions on AdS which induce logarithmic conformal field
theory (LCFT) on its boundary.
 Correlation functions in a
logarithmic conformal field theory exhibit logarithmic behavior
and first were noted by Gurarie[6]. Logarithmic operators appear
when two (or more ) operators are degenerate and have the same
dimension ,hence the Hamiltonian becomes non-diagonizable.In the
simplest case one has a pair $\hat{A}$ and $\hat{B}$ transforming
as:
 \begin{equation}
 \hat{A}(\lambda z)=\lambda^{-\Delta}\hat{A}(z)\nonumber
 \end{equation}
 \begin{equation}   \label{eq:a3}
\hat{B}(\lambda
z)=\lambda^{-\Delta}[\hat{B}(z)-\hat{A}(z)\ln\lambda]
\end{equation}
The bulk action which give rise to logarithmic operators on the
boundary were first described in[7,8]. A new method for
investigating LCFT is via nilpotent variables which was
introduced in [9] and then was modified in [10,11]. In reference
[11] a superfield was defined using a grassmanian variable $\eta$
and different components of a logarithmic pair and fermionic
fields.
\begin{equation}\label{eq:a4}
\hat{O}(\vec{x},\eta)=\hat{A}(\vec{x})+\hat{\bar{\zeta}}(\vec{x})\eta+\bar{\eta}\hat{\zeta}(\vec{x})+\bar{\eta}\eta\hat{B(\vec{x})}
 \end{equation}
  where $\hat{\zeta}(\vec{x})$and$
 \hat{\bar{\zeta}}(\vec{x})$ are fermionic
 fields with the same conformal dimension as $
 \hat{A}(\vec{x})$ and $ \bar{\eta}\eta $ acts
 as the nilpotent variable .
 Now it is easy to see that $\hat{O}(\vec{x},\eta)$ has the following
 transformation law:
  \begin{equation}\label{eq:a5}
  \hat{O}(\lambda\vec{x},\eta)=\lambda^{-(\Delta+\bar{\eta}\eta)}\hat{O}(\vec{x},\eta)
  \end{equation}
  If $ \hat{O}(\vec{x},\eta)$ were the logarithmic
  operator on the boundary of AdS the corresponding field
  $\Phi(x,\eta)$ on AdS can be extended as:
  \begin{equation}
  \Phi(x,\eta)=C(x)+\bar{\eta}\alpha(x)+\bar{\alpha}(x)\eta+
  \bar{\eta}\eta D(x)
  \end{equation}
  where x is (d+1)-dimensional coordinate with
  $x^{0},x^{1},...,x^{d}$ components and $
  x=(\vec{x},x^{d})$ and $x^{d}=0$ corresponds to the
  boundary.
  In reference [12] the following action was introduced for free
  scalar massive superfield with BRST symmetry.
  \begin{equation}
  S^{f}=-1/2\int d^{d+1}x\sqrt{|g|}\int d\bar{\eta}d\eta
  \nabla\Phi(x,\eta).\nabla\Phi(x,-\eta)+m^{2}(\eta)\Phi(x,\eta)\Phi(x,-\eta)
  \end{equation}

   where $m^{2}(\eta)=m_{1}^{2}+\bar{\eta}\eta m_{2}^{2}$ and $ m_{1}^{2}$,$m_{2}^{2}$ are defined to be$\Delta(\Delta-d)$
    and $(2\Delta-d)$ respectively and where g is the determinant of  the metric on
    AdS.

  Expanding the integrand of (8) in powers of $\eta$ and
  $\bar{\eta}$,integrating over $\eta$ and
  $\bar{\eta}$ and writing it in terms of four components one
  finds:
  \begin{eqnarray}
  S^{f}=-\frac{1}{2}\int d^{d+1}x\sqrt{|g|}[2\nabla C(x).\nabla
  D(x)+2m_{1}^{2}C(x)D(x)+m_{2}^{2}C^{2}(x)+   \nonumber\\
  2\nabla\bar{\alpha}(x).\nabla\alpha(x)+2m_{1}^{2}\bar{\alpha}(x)\alpha(x)]
  \end{eqnarray}
    We see that the bosonic part of this action is the same as the
  one proposed in [7].Also note that in the free theory, as we
  expect,the bosonic and fermionic part are decoupled.
  In the language of superfield the AdS/CFT correspondence becomes
  \begin{equation}
  \langle\exp\int d\bar{\eta}d\eta\int_{\partial AdS}d^{d}\vec{x}
     \hat{O}(\vec{x},\eta)\Phi_{b}(\vec{x},\eta)\rangle=\exp(-S_{cl}[\Phi_{b}(\vec{x},\eta)])
  \end{equation}
  where$ \Phi_{b}(\vec{x},\eta)$ is the value of $\Phi(x,\eta)$ on the
  boundary.
  In reference [12] the two-point correlation functions were
  calculated using equation (11),now our aim is to calculate
  n-point functions with this method.However the method used to
  calculate two-point functions will not give non-trivial 4-point
  functions unless interactions are also added.
  \section{Interactions and n-point functions}
 In order to find n-point functions[4,5,7,8,13] ($n\geq 3$)one should consider
 interaction terms in $\Phi(x,\eta)$ in addition to the free
 theory. Furthermore we wish the action to have BRST symmetry so that
 the correlation functions remain invariant under BRST
 transformations.In the language of superfield the
 infinitesimal BRST transformation is of the form:
 \begin{equation}\label{eq:a12}
 \delta\Phi(x,\eta)=(\bar{\epsilon}\eta+\bar{\eta\epsilon})\Phi(x,\eta)
 \end{equation}
 where $\bar{\epsilon}$ and $\epsilon$ are infinitesimal anticommuting
 parameters.
 If we consider the interaction term of degree n in
 $\Phi(x,\eta)$the general form is:
\begin{equation}
S^{I}=\lambda\int d^{d+1}x\sqrt{|g|}\int d\bar{\eta}d\eta
\end{equation}
\begin{equation}
V=\Phi(x,\eta_{1})...\Phi(x,\eta_{i})...\Phi(x,\eta_{n})
\end{equation}
with $\eta_{i}=n_{i}\eta$ and $n_{i}\in Z$.

 Then change of $V$ under BRST transformation is given by:
  \begin{equation}
 \delta V =\Phi(x,\eta_{1})...\Phi(x,\eta_{i})...\Phi(x,\eta_{n})[\bar{\epsilon}\sum_{i=1}^{n}\eta_{i}+\sum_{i=1}^{n}\bar{\eta}_{i}\epsilon]
 \end{equation}
 BRST invariance leads to:
 \begin{equation}
 \sum_{i=1}^{n}n_{i}=0
 \end{equation}
 which clearly does not have a unique solution.
 For n even the most symmetric choice is:
 \begin{equation}\label{eq:a18}
  V=\Phi^\frac{n}{2}(x,\eta)\Phi^\frac{n}{2}(x,-\eta)
  \end{equation}
 But for n odd we must choose an antisymmetric division such as:
\begin{equation}
V=\Phi^{n-1}(x,\eta)\Phi(x,-(n-1)\eta)\label{eq:a19}
\end{equation} which is
true for any n regardless of being odd or even.

 Let us first
consider interaction terms of type (\ref{eq:a18}) for n even.
\begin{equation}\label{eq:a20}
S^{I}=\int d^{d+1}x\sqrt{|g|}\int
d\bar{\eta}d\eta\frac{\lambda_{n}(\eta)}{n!}
\Phi^\frac{n}{2}(x,\eta)\Phi^\frac{n}{2}(x,-\eta)
\end{equation}
\begin{equation}
\lambda_{n}(\eta)=\lambda_{n}+\lambda_{n}'\bar{\eta}\eta
\end{equation}
 To write $S^{I}$ in terms of its four components,we
expand (\ref{eq:a20}) in powers of $\eta$'s,then integrate it
over them and we find:
\begin{equation}
\int d^{d+1}x\sqrt{|g|}[\frac{C^{n-1}(x)}{n!}
(\lambda'_{n}C(x)+n\lambda_{n}D(x))\frac{\lambda_{n}}{(n-1)!}
C^{n-2}(x)\bar{\alpha}(x)\alpha(x)]
\end{equation} It is
observed that the pure bosonic part of $S^{I}$is the same as the
on proposed in [7]and the fermionic part despite the free action
is coupled with bosonic field $C(x)$,this means that the exact
solution of equation of motion for this action and the one in [7]
will be different,but we will show that in tree level to first
order in $\lambda_{n}$ the correlation functions are the same.
 Now the total action is:
 \begin{equation}
 S=S^{f}+S^{I}
 \end{equation}
 The equation of motion for the the field $\Phi(x,\eta)$ is:
 \begin{equation}\label{eq:a24}
 (\nabla^{2}-m^{2})\Phi(x,\eta)+\frac{\lambda_{n}(\eta)}{(n-1)!}\Phi^{\frac{n}{2}}(x,\eta)\Phi^{\frac{n}{2}-1}(x,-\eta)=0
 \end{equation}
 Writing the equation of motion (\ref{eq:a24}) in terms of the four
 components we have:
 \begin{equation}\nabla^{2}C(x)-(m_{1})^{2}C(x)+\frac{\lambda_{n}}{(n-1)!}C^{n-1}(x)=0
\end{equation}
\begin{equation}
\nabla^{2}\alpha(x)-(m_{1})^{2}\alpha(x)+\frac{\lambda_{n}}{(n-1)!}C^{n-2}(x)\alpha(x)=0
\end{equation}
 \begin{equation}
 \nabla^{2}\bar{\alpha}(x)-(m_{1})^{2}\bar{\alpha}(x)+\frac{\lambda_{n}}{(n-1)!}C^{n-2}(x)\bar{\alpha}(x)=0
 \end{equation}
 \begin{eqnarray}\nabla^{2}D(x)-(m_{1})^{2}D(x)-(m_{2})^{2}C(x)+\frac{C^{n-2}(x)}{(n-1)!}((n-1)\lambda_{n}D(x)+\lambda'_{n}C(x))+\nonumber\\
   \frac{\lambda_{n}}{(n-2)!}C^{n-3}(x)\bar{\alpha}(x)\alpha(x)=0
\end{eqnarray}
 We see that as pointed out earlier the equation of motion for
 D(x) is different with the corresponding one in [7],also as we
 expected the fermionic fields to behave like ordinary field with
 dimension $\Delta$ interacting with C(x).
 The Dirichlet Green function for this system satisfies the
 equation:
   \begin{equation}
   (\nabla^{2}-m^{2})G(x,y,\eta)=\delta(x-y)
   \end{equation}
 together with the boundary condition:
 \begin{equation}
 G(x,y,\eta)|_{x\epsilon\partial AdS}=0
 \end{equation}
 The classical field $\Phi(x,\eta)$ satisfying equation of motion
 (24)with Dirichlet boundary condition on $\partial AdS$ satisfies
  the integral equation:
 \begin{eqnarray}
 \Phi(x,\eta)=\int_{\partial AdS}d^{d}y\sqrt{|h|}n^{\mu}\frac{\partial}{\partial
 y^{\mu}}G(x,y,\eta)\Phi_{b}(y,\eta)-\nonumber\\
 \int_{AdS}d^{d+1}y\sqrt{|g|}G(x,y,\eta))\frac{\lambda_{n}(\eta)}{(n-1)!}\Phi^{\frac{n}{2}}(x,\eta)\Phi^{\frac{n}{2}-1}(x,-\eta)=0
\end{eqnarray}
 where h is the determinant of the induced metric on $\partial AdS$
 and $n^{\mu}$ the unit vector normal to $\partial AdS$ and
 pointing outwards.
 We shall denote the surface term in (25)by $\Phi^{0}(x,\eta)$ and
 the remainder by $\Phi^{1}(x,\eta)$. Then substituting the
 classical solution (28)into (23) integrating by part and using
 the properties of Green function we obtain to tree level:
 \begin{equation}
 S_{cl}=-\frac{1}{2}\int d^{d}\vec{x}\sqrt{|h|}\int d\bar{\eta}d\eta n^{\mu}\Phi^{0}(x,\eta)
 \partial_{\mu}\Phi^{0}(x,\eta)+\int d^{d+1}x\sqrt{|g|}\int d\eta
 d\bar{\eta}\frac{\lambda_{n}(\eta)}
 {n!}(\Phi^{0}(x,\eta))^{n}
 \end{equation}
 The Green function for this problem is calculated in reference
 [12] substituting it for $\Phi^{0}(x,\eta)$ one obtains:
   \begin{equation}
\Phi^{0}(x,\eta)=a(\eta)\int
d^{d}\vec{y}(\frac{x^{d}}{(x^{d})^{2}+|\vec{x}-\vec{y}|^{2}})^{\Delta+\bar{\eta}\eta}\Phi_{b}(\vec{y},\eta)
\label{eq:a30}
\end{equation}
 with
 \begin{equation}
 a(\eta)=\frac{\Gamma(\Delta+\bar(\eta)\eta)}{2\pi^{d/2}\Gamma(\alpha+1)}=a+\bar{\eta}\eta a'
 \end{equation}
 Using the solution (\ref{eq:a30}),the classical action to first order in
 $\lambda_{n}$ becomes:
\begin{eqnarray}
 S_{cl}^{I}=\int d^{d+1}x\int d^{d}\vec{y}_{1}...d^{d}\vec{y}_{n}\int d\bar{\eta}d\eta
  a^{n}(\eta)\times\frac{(x^{d})^{-(d+1)+n(\Delta+\bar{\eta}\eta)}}{\prod_{i=1}^{n}[(x^{d})^{2}+|\vec{x}-\vec{{y}_{i}}|^{2}]^{\Delta+\bar{\eta}\eta}}\times \nonumber\\
 \Phi_{b}(\vec{y}_{1},\eta)...\Phi_{b}(\vec{y}_{n/2},\eta)\Phi_{b}(\vec{y}_{n/2+1},-\eta)...\Phi_{b}(\vec{y}_{n},-\eta)
\end{eqnarray}
 After expanding in powers of $\eta$ and $\bar{\eta}$ and
 integrating over $\eta$'s , one obtains for the classical
 solution:
 \begin{equation}
 S_{cl}^{I}=\frac{a^{n}}{n!}\int d^{d+1}x\int d^{d}\vec{y}_{1}...d^{d}\vec{y}_{n}[\lambda'_{n}\Psi_{1}+\lambda_{n}(\Psi_{2}+n\frac{a'}{a}+\ln\frac{(x^{d})^{n}}{\prod_{i=1}^{n}[(x^{d})^{2}+|\vec{x}-\vec{y_{i}}|^{2}]})]
 \end{equation}
with
\begin{equation}
J_{n}(\vec{y_{1}}....\vec{y_{n}},x)=\frac{(x^{d})^{-(d+1)+n\Delta}}{\prod_{i=1}^{n}[(x^{d})^{2}+|\vec{x}-\vec{y_{i}}|^{2}]^{\Delta}}
\end{equation} and
\begin{equation}
\Phi_{b}(\vec{y}_{1},\eta)...\Phi_{b}(\vec{y}_{n/2},\eta)\Phi_{b}(\vec{y}_{n/2+1},-\eta)...\Phi_{b}(\vec{y}_{n},\eta)=\Psi_{1}+\bar{\eta}\Psi+\bar{\Psi}\eta+\bar{\eta}\eta\Psi
\end{equation} Now we can derive the correlation of operator
fields on boundary by using equation (11).  Expanding both sides
of equation (11) in powers of $\Phi_{b}(\vec{x},\eta)$ and
integrating over $\eta$'s the n-point functions of different
components of $\hat{O}(\vec{x},\eta)$ can be found.So the
connected part of the tree level n-point function to order
$\lambda_{n}$ for components of $\hat{O}(\vec{x},\eta)$ are:
\begin{equation}
\langle\hat{B}(y_{1}....\hat{B}(y_{n})\rangle_{conn}=a^{n}\int
d^{d+1}x
J_{n}(\vec{y}_{1}\cdots\vec{y_{n}},x)\lambda'_{n}+\lambda_{n}(n\frac{a'}{a}+
\ln\frac{(x^{d})^{n}}{\prod_{i=1}^{n}[(x^{d})^{2}+|\vec{x}-\vec{y_{i}}|^{2}]})]
\end{equation}
\begin{equation}
\langle\hat{B}(y_{1})....\hat{B}(y_{i-1})\hat{A}(y_{i})\hat{B}(y_{i+1})....\hat{B}(y_{n})\rangle_{conn}=-\lambda_{n}a^{n}I_{n}(\vec{y_{1}}....\vec{y_{n}})
\end{equation}
\begin{equation}
\langle\hat{B}(y_{1})....\hat{B}(y_{i-1})\hat{\bar{\zeta}}(y_{i})\hat{B}(y_{i+1})...\hat{B}(y_{j})\hat{\zeta}(y_{j})\hat{B}(y_{j+1})...\hat{B}(y_{n})\rangle_{conn}=\pm\lambda_{n}a^{n}I_{n}(\vec{y_{1}}....\vec{y_{n}})
\end{equation}
 with
 \begin{equation}
I_{n}(\vec{y_{1}}....\vec{y_{n}})=\int d^{d+1}x
J_{n}(\vec{y_{1}}...\vec{y_{n}},x)
\end{equation}
where the minus sign in (35-3) refers to cases
$(i<j<n/2),(n/2<i<j)$ and $(i>n/2,j<n/2)$and all the other
correlation functions being zero. As we expected logarithmic
terms appear in the correlation functions of $\hat{B}$'s with
themselves and fermionic fields behave just like ordinary fields
of dimension $\Delta$,but their fermionic nature inhibits
appearing of more than two fermionic fields in nonzero correlation
functions.The integral $I_{n}$ can be made simpler after
integrating over $x^{d}$ and using Feynman parameterization [4].
The result for $(n=4)$ is:
\begin{equation}
I_{4}=\frac{2\Delta-\frac{d}{2}}{\Gamma(2\Delta)}\times\frac{2\pi^{\frac{d}{2}}}{(\eta\zeta\prod_{i<j}y_{ij})^{\frac{2}{3}\Delta}}\int_{0}^{\infty}F(\Delta,\Delta,2\Delta;1-\frac{(\eta+\zeta)^{2}}{(\eta\zeta)^{2}}-\frac{4}{\eta\zeta}\sinh^{2}(z))dz
\end{equation}
with
 \begin{equation}
y_{ij}=|\vec{y_{i}}-\vec{y_{j}}|
\eta=\frac{y_{12}y_{34}}{y_{14}y_{23}}
\zeta=\frac{y_{12}y_{34}}{y_{13}y_{24}}
\end{equation}
Coming back to odd powers of $\Phi(x,\eta)$, choosing it of the
form (19) we observe that it is not possible to write a consistent
equation of motion,for $\Phi(x,\eta)$ as a whole. However
integrating over $\eta$'s ,one can derive a consistent set of
equations for the components. Our method does have the weakness
that choosing either of the forms (18) and (19) is arbitrary. Even
requiring BRST symmetry does not fix the choice. It is not clear
to the authors what extra requirement is necessary to fix this
choice. When the power is even, of course an extra symmetry under
reflection   $\Phi(x,\eta)\rightarrow  -\Phi(x,\eta)$ exists!
Therefore only in the case of even powers, a choice can be fixed.

\end{document}